\documentclass[a4paper,fleqn,usenatbib]{mnras}

\usepackage{newtxtext,newtxmath}

\usepackage[T1]{fontenc}
\usepackage{ae,aecompl}

\usepackage{epsfig}
\usepackage{epstopdf}

\usepackage{graphicx}	
\usepackage{amsmath}	
\usepackage{amssymb}	
\usepackage{float}
\usepackage{longtable}
\usepackage{sidecap}
\usepackage{hyperref}
\usepackage{threeparttable}
\usepackage{booktabs,caption}
\include{def_bibtex}

\title[The X-rays plateau phase of 4U 0115+63]{The low luminosity behaviour of the 4U 0115+63 Be/X-ray transient}

\author[A. Rouco Escorial et al.]{
\Large A. Rouco Escorial,$^{1}$\thanks{E-mail: A.RoucoEscorial@uva.nl (ARE)}
A.S. Bak Nielsen,$^{2}$
R. Wijnands,$^{1}$
Y. Cavecchi,$^{3,4}$
N. Degenaar,$^{1}$
A. Patruno$^{2,5}$
\\
$^{1}$Anton Pannekoek Institute for Astronomy, University of Amsterdam, Science Park 904, 1098 XH, Amsterdam, The Netherlands\\
$^{2}$Leiden Observatory, Leiden University, Neils Bohrweg 2, 2333 CA, Leiden, The Netherlands\\
$^{3}$Department of Astrophysical Sciences, Princeton University, Peyton Hall, Princeton, NJ 08544, USA\\
$^{4}$Mathematical Sciences and STAG Research Centre, University of Southampton, SO17 1BJ, UK\\
$^{5}$ASTRON, the Netherlands Institute for Radio Astronomy, Postbus 2, 7900 AA, Dwingeloo, The Netherlands\\
}

\date{Accepted 2017 August 11. Received 2017 August 11; in original form 2017 April 2}

\pubyear{2017}

\begin{document}
\label{firstpage}
\pagerange{\pageref{firstpage}--\pageref{lastpage}}
\maketitle

\begin{abstract}
The Be/X-ray transient 4U 0115+63 exhibited a giant, type-II outburst in October 2015. The source did not decay to its quiescent state but settled in a meta-stable plateau state (a factor $\sim$10 brighter than quiescence) in which its luminosity slowly decayed. We used \textit{XMM-Newton} to observe the system during this phase and we found that its spectrum can be well described using a black-body model with a small emitting radius. This suggests emission from hot spots on the surface, which is confirmed by the detection of pulsations. In addition, we obtained a relatively long ($\sim$7.9 ksec) \textit{Swift}/XRT observation $\sim$35 days after our \textit{XMM-Newton} one. We found that the source luminosity was significantly higher and, although the spectrum could be fitted with a blackbody model the temperature was higher and the emitting radius smaller. Several weeks later the system started a sequence of type-I accretion outbursts. In between those outbursts, the source was marginally detected with a luminosity consistent with its quiescent level. We discuss our results in the context of the three proposed scenarios (accretion down to the magnestospheric boundary, direct accretion onto neutron star magnetic poles or cooling of the neutron star crust) to explain the plateau phase.\\
\end{abstract}


\section{Introduction}\label{sec:introduction}
Be/X-ray transients harbour high-magnetic field (B$\sim$10$^{12-13}$ G) neutron stars (NSs) that move around Be-type stars in eccentric orbits. Typically, these binaries can exhibit two types of transient activity: more common, short-lived (i.e., small fraction of the orbital period) type-I outbursts that typically occur at the periastron passage of the NSs, and giant, type-II outbursts that can last significantly longer than an orbital period. In the first case, the NSs accrete matter from the Be-star decretion disks (e.g., \citealt{Okazaki2001}) and reach X-ray luminosities of L$_{X}\sim$10$^{36-37}$ erg s$^{-1}$. In the second case, during the type-II outbursts the luminosities usually reach the Eddington limit for a NS but the exact physical mechanism(s) causing these outbursts is not understood (\citealt{Moritani2013}; \citealt{Monageng2017}).\\
\indent Most studies of Be/X-ray transients have been performed at high luminosities \citep[i.e., $>$10$^{36}$ erg s$^{-1}$; e.g., see the review by][]{Reig2011} and consequently not much is known about their behaviour when they accrete at lower X-ray luminosity. However, several Galactic systems have been detected at luminosities of L$_{X}\sim$10$^{34-35}$ erg s$^{-1}$ (e.g., \citealt{Motch1991}; \citealt{Campana2002}; \citealt{Rutledge2007}) indicating that they are still accreting but at relatively low rates. In addition, detailed X-ray studies (using the \textit{Chandra} and \textit{XMM-Newton} observatories) of the Be/X-ray binary populations in the Magellanic Clouds (i.e., the Small Magellanic Cloud; \citealt{Laycock2010}; \citealt{Haberl2016}) have showed that a significant number of the Be/X-ray transients in those galaxies, can be detected at similar low luminosities in between their outbursts. This fact indicates that such low-luminosity states are a common property of Be/X-ray transients.\\
\indent At these low rates, the NS spin period is a key component in determining how the accretion proceeds. For slow spinning NSs (spin periods of hundreds of seconds) matter may still be directly accreted onto the stellar surface (i.e., at the magnetic poles) but for the fast spinning systems (with spin periods of only a few seconds) matter is likely ejected from the inner part of the system due to the pressure of the rotating NS magnetic field (the so-called propeller effect; e.g., \citealt{Illarionov1975}; \citealt{Stella1986}; \citealt{Romanova2004}; \citealt{DAngelo2010}; \citealt{Tsygankov2016}). In the latter case, one would expect that the observed flux is not pulsed. However some sources still exhibit pulsations in this regime, probably because matter is able to leak through the field lines and reach the surface at the magnetic poles, thus creating hot spots (e.g., \citealt{Elsner1977}; \citealt{Ikhsanov2001}; \citealt{Lii2014}).\\
\indent Some sources are detected at even lower luminosities of L$_{X}\sim$10$^{32-34}$ erg s$^{-1}$ (e.g., \citealt{Mereghetti1987}; \citealt{Roberts2001}; \citealt{Campana2002}; \citealt{Reig2014}; \citealt{Elshamouty2016}), but the cause of this faint emission is still unclear. One possibility is the accretion down to the magnetosphere although it is likely that this emission would not peak in the X-rays but at longer wavelengths (e.g., UV; \citealt{Tsygankov2016}). However, the pulsed emission seen in some systems at those luminosities (e.g., \citealt{Rothschild2013}; \citealt{ Doroshenko2014}; Table 2 of \citealt{Reig2014}) suggests that, for at least those systems, the matter does reach the surface at the magnetic poles, supporting the idea of leakage of matter through the magnetospheric barrier (\citealt{Orlandini2004}; \citealt{Mukherjee2005}).\\
\indent Although low level accretion onto the magnetic poles could be the physical mechanism behind these low luminosities in some sources, it is also possible that in other systems the accretion has fully halted and we see a cooling NS. During outburst the accreted matter might heat up the NS (due to the pycnonuclear reactions deep in the crust; \citealt{Brown1998}) and, when the accretion has stopped, the deposited heat is radiated away. This explanation has been proposed for several sources but it remains to be confirmed  (\citealt{Campana2002}; \citealt{Wijnands2013}; \citealt{Reig2014}; \citealt{Elshamouty2016}; \citealt{Tsygankov2017b}).\\
\indent In the heating and cooling scenario, one would expect that after a long and strong period of accretion (i.e., after type-II outbursts), the crust might be heated very significantly and it might have become hotter than the core. When the accretion has stopped, the crust would then slowly cool down until equilibrium is reached again. This process would be similar to what has been observed for several low-magnetic field NS systems (for recent discussions see \citealt{Degenaar2015} and \citealt{Parikh2017}). \cite{Wijnands2016} attempted to test this hypothesis in two Be/X-ray transients (4U 0115+63 and V0332+53) after the end of their type-II outbursts. They found that after those bright outbursts both sources showed elevated emission in a meta-stable state above their known quiescent levels (e.g., \citealt{Elshamouty2016}; \citealt{Tsygankov2017b}). \cite{Wijnands2016} suggested that this "plateau phase" could indeed be consistent with the slow cooling of the crust, although they could not exclude accretion scenarios. Here we further investigate this plateau phase for 4U 0115+63.\\

\section{Observations, Analysis and Results}\label{sec:analysis}
After the \textit{Swift}/X-Ray Telescope (XRT) data reported in \cite{Wijnands2016}, we obtained several extra \textit{Swift}/XRT observations as well as an \textit{XMM-Newton} one of 4U 0115+63 (Table \ref{tab:Table1} and Table \ref{tab:Table2}). This source harbours a magnetized NS (B$\sim$1.3$\times$10$^{12}$ G; \citealt{Raguzova2005}), with a spin period of P$_\mathrm{s}\sim$3.62 s (\citealt{Cominsky1978}) and an orbital period of P$_\mathrm{orb}\sim$24.3 days (\citealt{Rappaport1978}). 4U 0115+63 was monitored with the XRT in the Photon Counting (PC) mode and observed by \textit{XMM-Newton} with the EPIC detectors in the full frame (pn) and large window (MOS) modes. The \textit{Swift}/XRT PC mode only allows for a time resolution of $\sim$2.5 s which is insufficient to study possible pulsations due to the NS spin. For \textit{XMM-Newton}, we used the pn to search for and study the pulsations (time resolution 73.4 ms; section \ref{subsec:timing}) due to its higher count rate.\\
\begin{figure}
\centering
\includegraphics[width=1\columnwidth, trim=1cm 0 1cm 2.5cm]{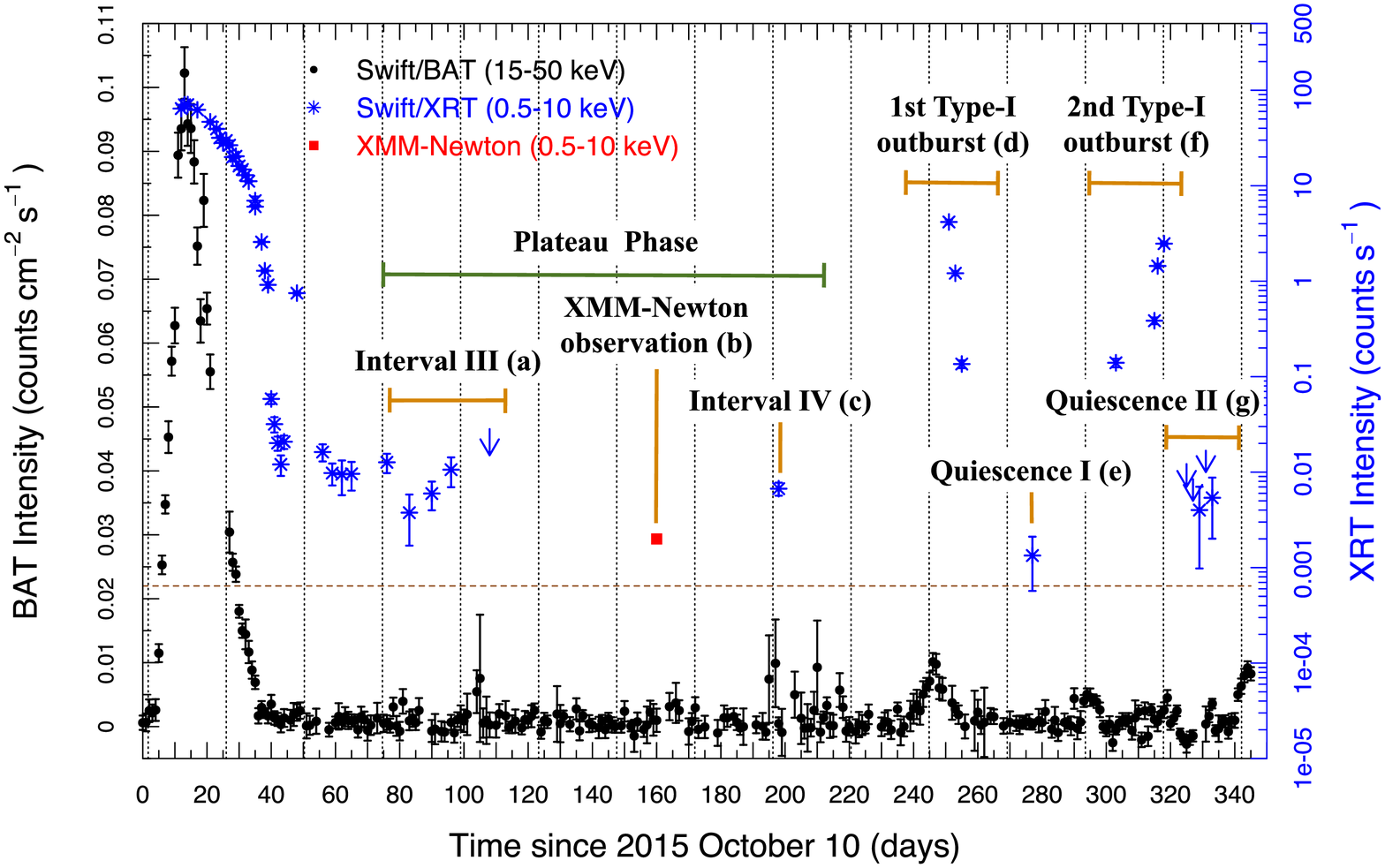}
    \caption{\textit{Swift}/XRT (blue) and \textit{Swift}/BAT (black) light curves during and after the 2015 type-II outburst of 4U 0115+63. The time of our \textit{XMM-Newton} observation is given as a red symbol converted in \textit{Swift}/XRT count rate (section \ref{subsec:lightcurve}). The dotted brown line corresponds to the quiescent level of 4U 0115+63. Vertical dotted lines indicate the time of periastron passages. We label our results in the different source phases as a continuation of the ones used in \citet*{Wijnands2016}. Errors are 3$\sigma$.}
   \label{fig:BAT_lc}
\end{figure}

\subsection{Light curve}\label{subsec:lightcurve}
The 2015 type-II outburst was monitored using the \textit{Swift}/BAT and \textit{Swift}/XRT (see Figure \ref{fig:BAT_lc}). In particular, the decay of this outburst was intensively monitored using the \textit{Swift}/XRT which also allowed to observe the transition of the source into the low-luminosity state (see also \citealt{Wijnands2016}). This was the first time such a state was detected for this system but very likely it was present after the end of the previous type-II outbursts as well. However, those possible occurrences were missed because no sensitive X-ray observations after the end of those outbursts were obtained in the past.\\
\indent We obtained the \textit{Swift}/BAT data from the hard X-ray transient monitor web page\footnote{\url{http://swift.gsfc.nasa.gov/results/transients/weak/4U0115p634/}} (\citealt{Krimm2013}) and the \textit{Swift}/XRT light curve from the interface build \textit{Swift}/XRT products\footnote{\url{http://www.swift.ac.uk/user_objects/}} (\citealt{Evans2009}). For a direct comparison the source count rate during the \textit{XMM-Newton} observation was converted to XRT count rate (for the energy range 0.5-10 keV) applying the WEBPIMMS\footnote{\url{http://heasarc.gsfc.nasa.gov/cgi-bin/Tools/w3pimms/w3pimms.pl}} tool and the spectral parameters obtained from the first observation during the low luminosity state (hereafter "plateau phase") in our paper. From Figure \ref{fig:BAT_lc} (see also \citealt{Wijnands2016}), we can see that after the type-II outburst the source did not go directly to quiescence (\citealt{Campana2002}; \citealt{Tsygankov2017b}) but settled down at a plateau that is a factor of $\sim$10 brighter. Almost $\sim$240 days after the type-II outburst, the source experienced several type-I outbursts (e.g., \citealt{Nakajima2016a,Nakajima2016b,Nakajima2016c}).\\
\indent After the episodes reported by \cite{Wijnands2016}, the \textit{Swift}/XRT count rate continued to decrease until our "\textit{XMM-Newton} observation (b)" (see Figure \ref{fig:BAT_lc}): from $\sim$0.015 counts s$^{-1}$ during "Interval I" of \citet*[i.e., at the start of the plateau phase]{Wijnands2016} to $\sim$0.002 counts s$^{-1}$ during \textit{XMM-Newton} observation (b). A similar decreasing trend is seen in the X-ray luminosities (Figure \ref{fig:Parameters}, top panel) obtained from our spectral analysis (see section \ref{subsec:spectra}). Despite that a general decreasing trend is seen in the \textit{Swift}/XRT count rate, it unexpectedly increased slightly again to $\sim$0.007 count s$^{-1}$ ("Interval IV (c)") after the\textit{ XMM-Newton} observation (b). During the plateau phase the \textit{Swift}/BAT count rate appears to have brief excursions above the background possible indicating occasional enhanced activity (Figure \ref{fig:BAT_lc}). However, those variations are consistent with statistical fluctuations and do not represent detections of the source (see \citealt{Krimm2013} for a detailed description about the \textit{Swift}/BAT data processing). The monitoring campaign stopped for $\sim$53 days after our Interval IV (c) \textit{Swift}/XRT observation until the source exhibited a type-I outburst ("1$^{st}$ Type-I outburst (d)") during which the \textit{Swift}/XRT count rate reached $\sim$4 counts s$^{-1}$. The decay during this outburst was monitored with the \textit{Swift}/XRT and 26 days after the peak of the outburst we obtained a 8.8 ksec observation in which the source only showed 8 photons ("Quiescence I (e)"; see Figure \ref{fig:BAT_lc}). This level is consistent with the quiescent level of the source (Figure \ref{fig:BAT_lc}; \citealt{Tsygankov2017b}).\\
\indent The \textit{Swift}/BAT continued to monitor the source and 17 days later another type-I outburst was detected (Figure \ref{fig:BAT_lc}; \citealt{Nakajima2016b}). We monitored the source using the XRT to determine how it would transit to quiescence. Surprisingly during the first observation the source was not in quiescence but at relatively high count rate ($\sim$0.15 counts s$^{-1}$) and 12 days later the source increased again in count rate indicating the start of the next type-I outburst ("2$^{nd}$ Type-I outburst (f)"; Figure \ref{fig:BAT_lc}). Already within 7 days of the peak , the source was back in quiescence; we obtained 5 observations ("Quiescence II (g)") during which the source was not detected or only marginally. The stacking of  these observations shows a weak (9 photons in 4.7 ksec) but clear detection.\\

\subsection{Spectral analysis}\label{subsec:spectra}
\begin{figure}
\centering
\includegraphics[width=\columnwidth, trim=0.5cm 0 1.25cm 0]{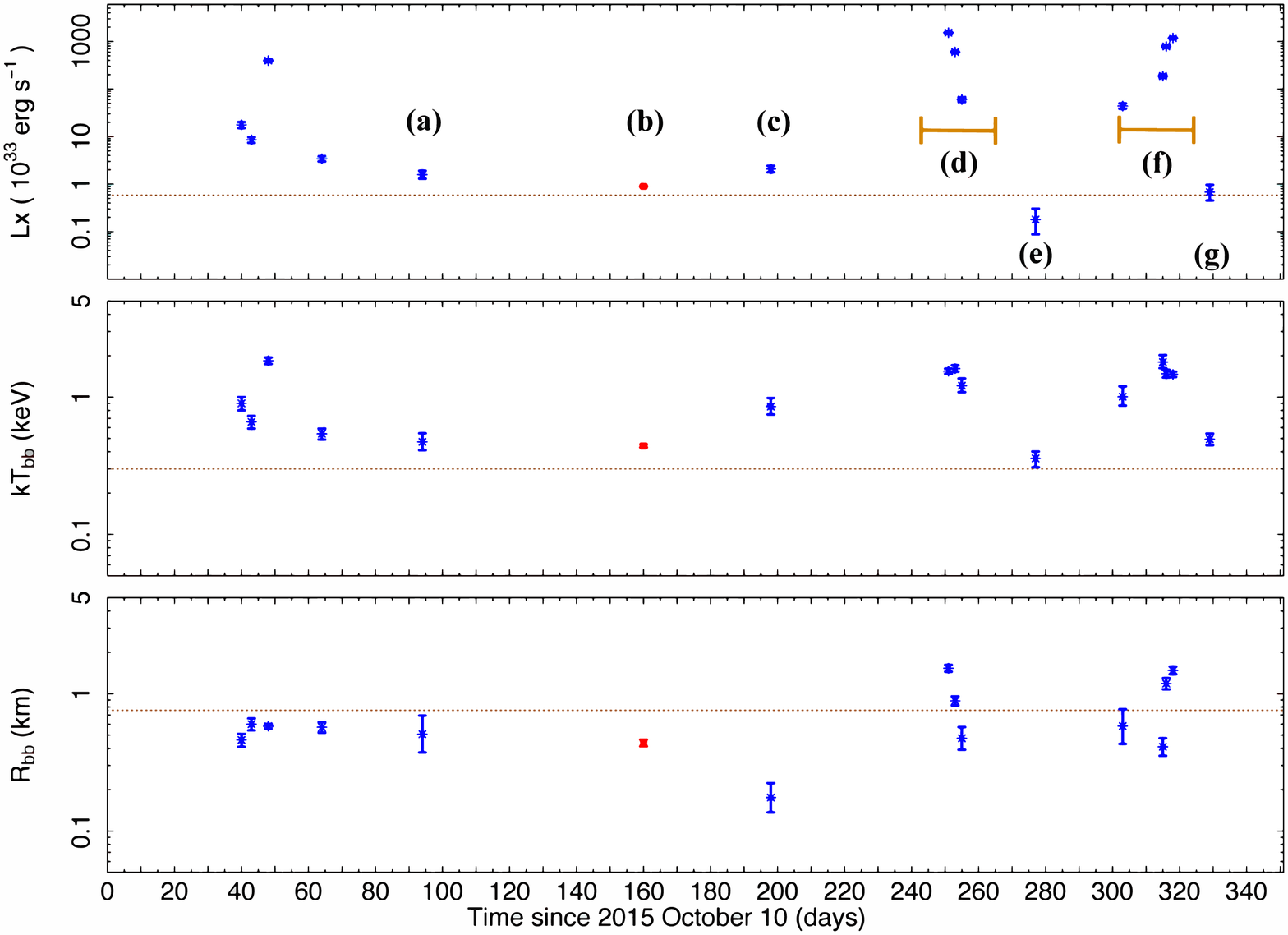}
    \caption{Evolution (using a black-body model) of the luminosity (top; 0.5-10 keV), temperature (middle) and emission radius (bottom; fixed values not shown). The blue points are the \textit{Swift}/XRT observations and the red point is the \textit{XMM-Newton} one. The dotted brown line corresponds to parameters in quiescence (\citealt{Tsygankov2017b}). Labels as in Figure \ref{fig:BAT_lc}. Errors are 1$\sigma$.}
   \label{fig:Parameters}
\end{figure}
For the spectral analysis, we only used \textit{Swift}/XRT observations obtained after those already analyzed by \cite{Wijnands2016} (ObsID 000311720[42-59]) and used the results obtained by these authors for the earlier observations. The data extraction and analysis have been performed in the same way as in \cite{Wijnands2016} so that the results can be directly compared. Due to the very low count rates in the observations with ObsIDs 000311720[42-45] (a) and 000311720[55-59] (g), we stacked those ones. The data were analysed with HEASOFT v.6.17. After reprocessing the raw data with the XRTPIPELINE, we extracted source and background spectra using XSELECT. The source information (count rates, spectra) was obtained from a circular region with a radius of 15 pixels centered around the source position (\citealt{Reig2015}), and background information from a surrounding annulus with an inner-outer radius of 60-110 pixels. Only during the type-I outbursts observations the source count rate was $\geq$0.5 counts s$^{-1}$ causing piled-up. Therefore, those data were corrected following the standard thread\footnote{\url{http://www.swift.ac.uk/analysis/xrt/pileup.php}}. The exposure maps were obtained through XRTPIPELINE and were used to create ancillary response files using XRTMKARF. We employed the response matrix files (version 14) from the \textit{Swift} calibration database.\\
\indent For the \textit{XMM-Newton} data, we used SAS (version 1.2) for reducing and analysing them. The EMPROC and EPPROC tasks produced calibrated event lists, which were filtered against background flaring (determined in the 10-12 keV range with a threshold rate$\geq$0.5 counts s$^{-1}$ for pn; in the $>$10 keV range with a threshold rate$\geq$0.3 counts s$^{-1}$ for MOS). Source counts and spectra were obtained from a circular region with a 20 arcsec radius, and the background ones from a circular region free of sources with a 50 arcsec radius on the same CCD. The data were not piled-up. The response matrix files were generated using RMFGEN and the ancillary response files by ARFGEN.\\
\indent In Figure \ref{fig:Parameters} we show the evolution of the source during the plateau phase plotting our new results with those of \cite{Wijnands2016}. Due to the often very few counts in the spectra we rebinned the data (with GRPPHA for \textit{Swift} and SPECGROUP for \textit{XMM-Newton} data) to 1 count per bin and we used C-statistics to fit the spectra. The data were fitted in the 0.5-10 keV energy range using XSPEC v.12.9.0. The pn, MOS1 and MOS2 spectra were fitted simultaneously, with all parameters tied between the spectra. We fitted the data with either an absorbed power-law model (PEGPWRLW) or an absorbed blackbody model (BBODYRAD). We note that we cannot exclude other single component models or spectral models with multiple components (e.g., a blackbody model plus a power-law contribution; potentially describing multiple emission mechanisms). However, the quality of our spectra is typically rather poor and we cannot distinguish between different single component model (except for the \textit{XMM-Netwon} spectrum, see below), let alone that we can obtain constraining results when fitting multiple components. Therefore, we have opted to only report on the spectral results obtained with two of the most basic models used in the fitting of X-ray binary spectra: the blackbody model and the power-law model. This will allow us to determine general trends in the data (i.e., softening of the spectra in time) which will help in the interpretation of the results.\\ \indent We included absorption by the interstellar medium (TBABS) with abundances set to WILM (\citealt{Wilms2000}) and cross-sections to VERN (\citealt{Verner1996}). The column density was fixed to the same value as in \citet[i.e., N$_{H}$=9$\times$10$^{21}$ cm$^{-2}$]{Wijnands2016}. We adopted a distance of D=7 kpc (\citealt{Negueruela2001}). In the PEGPWRLW model we set the energy boundaries to 0.5 and 10 keV, so that the model normalization gives the unabsorbed flux for that energy range (see Table \ref{tab:Table2}). For the BBODYRAD fits, we left the emitting radius as a free parameter and determined the unabsorbed 0.5-10 keV flux by using CFLUX (see Table \ref{tab:Table1}).\\
\indent The results of our spectral analysis are given in Tables \ref{tab:Table1} and \ref{tab:Table2}. Although we cannot statistically prefer one of the models over the other during the plateau phase because of the \textit{Swift} data quality, the \textit{XMM-Newton} spectrum is of high enough one that the black-body model is preferred. The PEGPWRLW fits suggest that the spectra of the plateau phase are softer (typical photon indices of 1.5-3.6) than those of the type-I outbursts (indices of 0.5-1.2; Table \ref{tab:Table2}). The spectra of the plateau phase can be adequately described by a BBODYRAD model with temperatures of kT$_{bb}\sim$ 0.4-0.8 keV with radii R$_{bb}\sim$ 0.1-0.5 km (Table \ref{tab:Table1}; Figure \ref{fig:Parameters}), smaller than the radius of a NS. This suggests that the emission comes from hot spots on the surface (likely at the magnetic poles). This is confirmed by the detection of pulsations in the \textit{XMM-Newton} data (section \ref{subsec:timing}). The (c) state shows a higher luminosity (L${_X}\sim$2.1$\times$10$^{33}$ erg s$^{-1}$; Figure \ref{fig:Parameters}) than during (b), as well as a higher temperature (from $\sim$0.44 keV to $\sim$0.85 keV) but a significantly smaller emission radius (from $\sim$0.51 to $\sim$0.18 km; Table \ref{tab:Table1}). The source is detected at L${_X}\sim$10$^{32-33}$ erg s$^{-1}$ (0.5-10 keV) during the low luminosity state, whereas during the type-I outbursts it increases to L${_X}\sim$10$^{34-36}$ erg s$^{-1}$. The (e) and (g) states are consistent with quiescence (Figure \ref{fig:Parameters}).\\

\subsection{Timing analysis}\label{subsec:timing}
\begin{figure}
\centering
\includegraphics[width=0.9\columnwidth, trim=2.5cm 1cm 4cm 2.5cm]{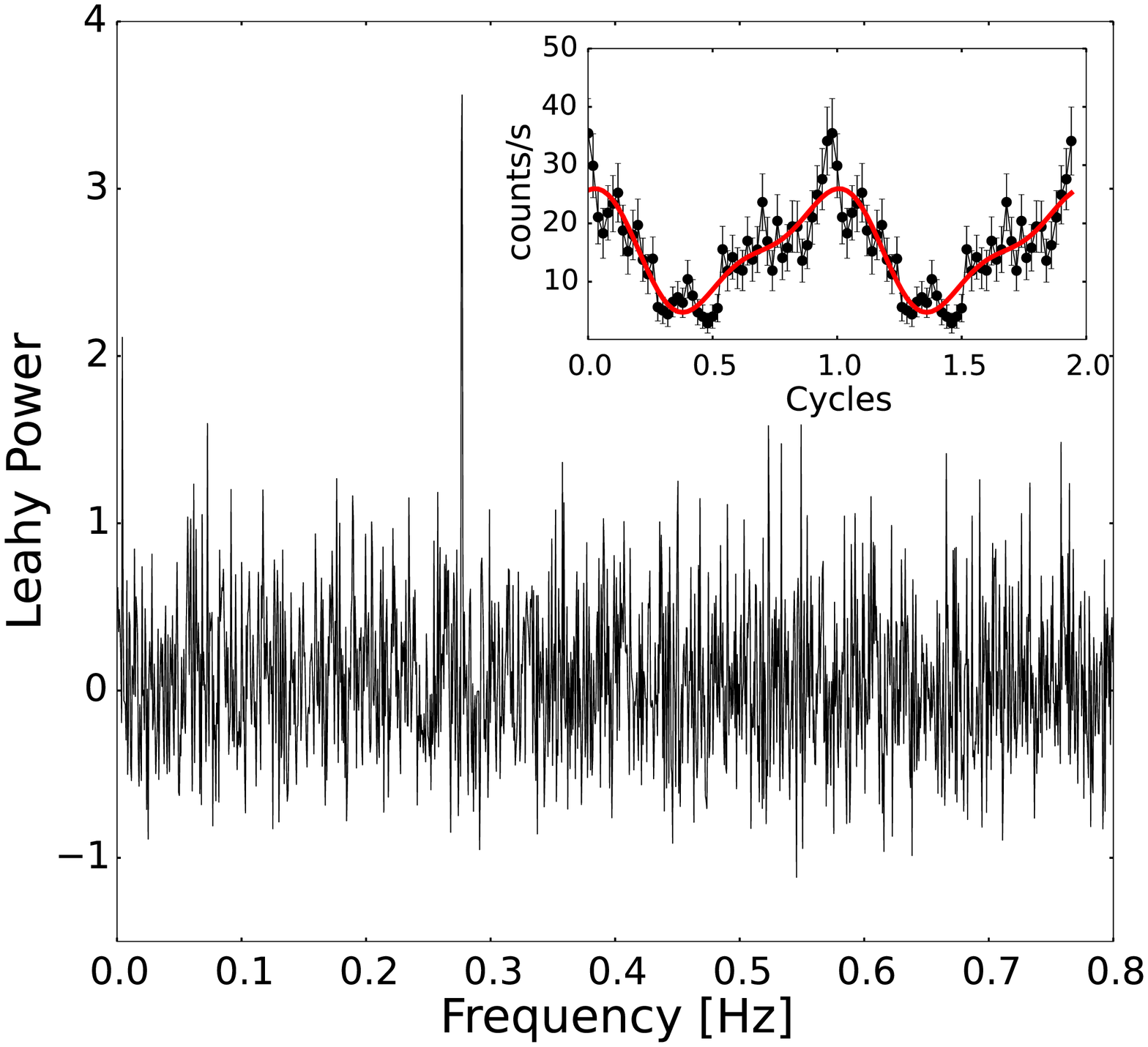}
    \caption{Power spectrum obtained from the EPIC-pn data showing a clear signal at $\sim$0.277 Hz that corresponds to the NS spin frequency. Inset: folded (using the detected frequency) pulse profile. Two cycles are shown for clarity. The red curve is a fit to the profile using two harmonically related sinusoids.}
   \label{fig:Pulse}
\end{figure}

The  \textit{XMM-Newton} pn data were barycentric corrected using the task BARYCEN and the region selection applied was the same as used in the spectral analysis. We rebinned the data to 0.1 s time resolution and then used 16384 points (thus 1638.4 s of data) to make a FFT. This resulted in a power-density spectrum in the frequency range of 6.104$\times$10$^{-4}$ Hz to 5 Hz (the Nyquist frequency). No background subtraction was performed prior to the calculation of the FFTs. The Poisson level was removed from the final power spectrum, which was then normalized using an rms normalization, where the power density units are (rms/mean)$^2$ Hz$^{-1}$. The power spectrum in Figure \ref{fig:Pulse} shows a peak close to the known NS spin frequency ($\sim$0.3 Hz).\\
\indent The light curve was then folded using PRESTO (\citealt{Ransom2001}) to determine the precise spin frequency. The best results were obtained using a frequency of (27677600.0$\pm$9.2)$\times$10$^{-7}$ Hz yielding a period of (361300.0$\pm$1.5)$\times$10$^{-5}$ s. The pulse profile (see inset in Figure \ref{fig:Pulse}) was fitted with a sinusoid using the procedure described in \cite{Patruno2010} to obtain the pulse fractional amplitude. We fitted one harmonic and two harmonics (to only one cycle but for clarity two cycles are shown in Figure \ref{fig:Pulse}). When one harmonic is fitted we find the fractional amplitude to be FA$_1$=83.3$\pm$9.5 per cent, with a $\chi^{2}_{\nu}$=1.3 for 46 degrees of freedom. This indicates that the fit is marginally acceptable with a p-value of 7$\%$. To explore the option that the profile might contain multiple harmonics, we fitted the profile with two harmonics. We found FA$_1$=83.3$\pm$8.7 and FA$_2$=28.4$\pm$7.6 per cent with $\chi^{2}_{\nu}$=1.09 for 44 degrees of freedom. Therefore, fitting two harmonics to the profile only yielded a marginally better $\chi^{2}_{\nu}$ and consequently we cannot conclude that the second harmonic is real. If it would be real, it might be due to slight beaming of the emission (e.g., by Compton scattering due to low-level accretion) or because there are two hot spots contributing to the observed emission. However, because of the low significance of this second harmonic we will not discuss it or its potentially consequences on the interpretation further in our paper.\\

\section{Discussion}\label{sec:discussion}
We have presented additional \textit{Swift}/XRT and \textit{XMM-Newton} observations (Figure \ref{fig:BAT_lc}) during the low-luminosity plateau phase previously identified for 4U 0115+63 (\citealt{Wijnands2016}). From our \textit{Swift} data we can infer that the plateau phase lasted between 160 and 200 days. The exact duration cannot be estimated due to the lack of monitoring during the final state of this state and because the system started to exhibit type-I outbursts. \cite{Wijnands2016} proposed three models to explain the plateau phase: accretion down to the NS magnetosphere, cooling of an accretion heated NS crust and direct low-level accretion onto the NS magnetic poles. The magnetospheric accretion model can likely be excluded because no pulsations are expected in this scenario (\citealt{Campana2002}), contrary to our \textit{XMM-Newton} results where we detect a clear pulsation at the NS spin frequency. This assumes that pulsations were also always present during the XRT observations in the plateau phase. The data do not allow this to be tested, but if they were not present during (some of) these XRT observations, it might still be possible that magnetospheric accretion could still have happened during some of those observations. Moreover, it is indeed possible that multiple emission mechanisms were active during different observations in the plateau phase (or even during a specific observation; e.g., cooling emission during some observations and accretion emission during others; i.e., at periastron), but the current data set is of too low quality to further investigate this. In the following two sections we, for simplicity, assume that only one emission mechanisms is active during the full duration of the plateau phase. This will allow us to investigate if indeed one single mechanism is sufficient to explain this enigmatic state and what the possible problems are with this hypothesis.\\

\subsection{Cooling of the neutron star heated crust}\label{subsec:cooling}
The accretion-induced heating of the NS crust and its subsequent cooling has been studied extensively for accreting low-magnetic field (B$\leq$10$^8$ G) NSs, but not for high magnetic-field NSs (B$\sim$10$^{12-13}$ G). The strong magnetic field could alter the heating and cooling processes significantly. \cite{Wijnands2016} suggest that the magnetic field configuration in the crust might be such that the cooling process could take place through hot spots, an idea supported by the small radii of the emission regions found in our spectral analysis. Such interpretation might be consistent with the detection of pulsations during our \textit{XMM-Newton} observation.\\
\indent Our \textit{Swift}/XRT monitoring campaign during the plateau phase is consistent with the cooling hypothesis, except for the last observation (Interval IV in Figure \ref{fig:BAT_lc}). During that observation the luminosity increases compared to the previous \textit{XMM-Newton} one, concurrently, we detect a significant increase in the fit temperature (Figure \ref{fig:Parameters}; Table \ref{tab:Table1}). The emitting region notably decreased in size but this was not enough to compensate the rise in temperature so that the observed luminosity increased. This indicates that the energy release at the hot spots increases significantly and this cannot straightforwardly be explained in the cooling hypothesis.\\
\indent Interval IV was obtained after a periastron passage which could lead to matter transfer from the Be disk or from the wind of the companion to the NS. This matter could leak through the magnetosphere barrier and potentially cause an increase in luminosity. The lack of monitoring after Interval IV does not allow us to determine if this observation was part of an overall trend or a special event on top of a general decay curve. Nevertheless, the spectral shape is difficult to explain in this scenario (section \ref{subsec:accretion}) and therefore we suggest an explanation for this enigmatic behaviour in the cooling scenario.\\
\begin{figure}
\centering
\includegraphics[width=1.2\columnwidth, trim=5cm 1cm 2cm 1.2cm]{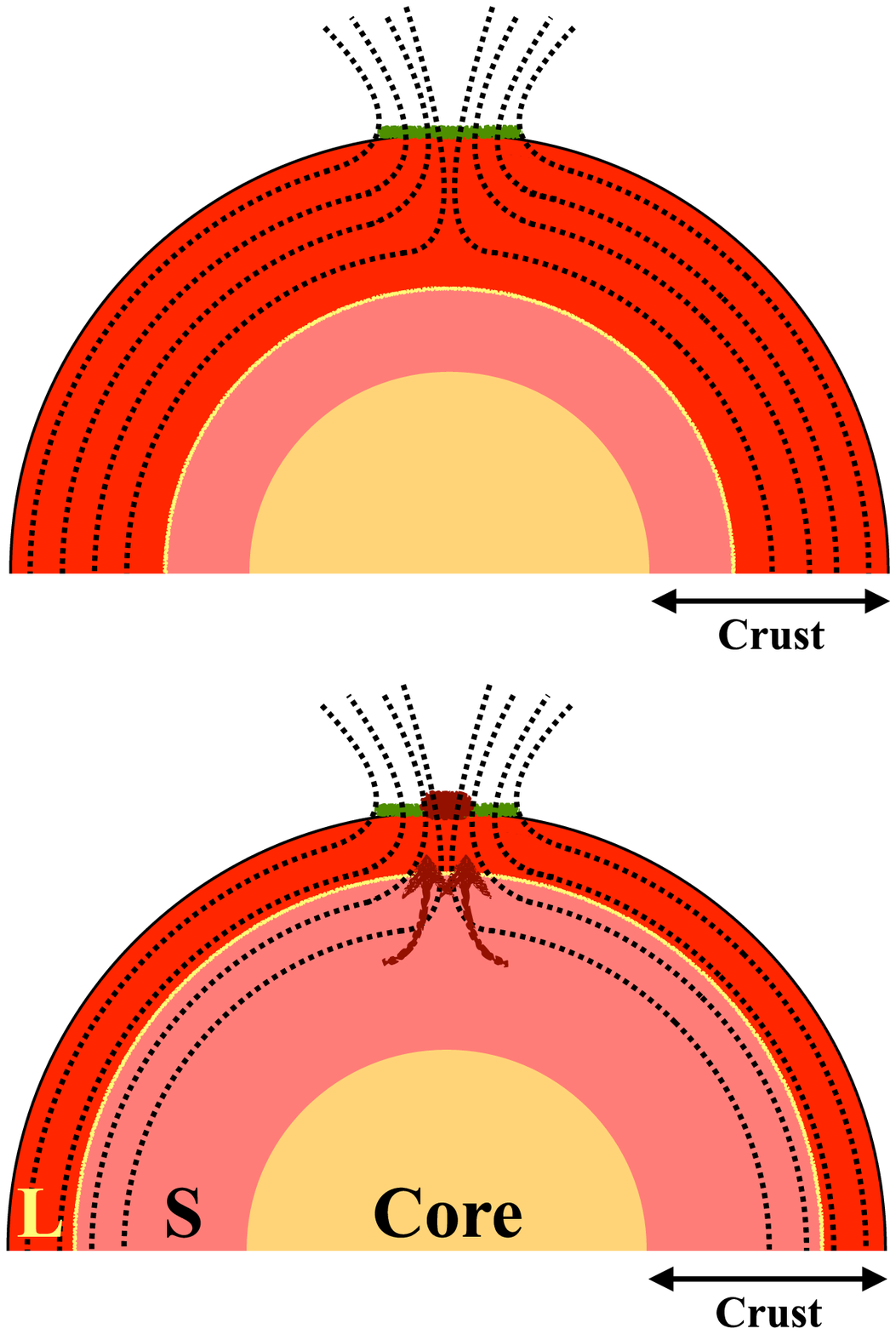}
    \caption{Phase Transition model (not to scale). Colour code: red (liquid crustal layer, so-called "ocean"), pink (solid crustal layer), yellow (core), green (hot spot region at the magnetic pole), garnet (hot spot central emission region) and dashed black lines (magnetic field lines). \textit{Upper Panel}: Initial state configuration of the crust once the accretion has halted. \textit{Bottom Panel}: Configuration after the cooling has proceeded. The inner layers of the ocean have become solid and the magnetic field lines that are placed in this region conduct the heat faster towards the magnetic pole. As a consequence, the central region of the hot spot emits higher heat flux.}
   \label{fig:Phase_model}
\end{figure}
\indent One possible explanation, which we call the "Phase Transition model" (Figure \ref{fig:Phase_model}), is based on the possible phase transitions the crust matter might undergo during the cooling process. Two matter states likely are present: a solid state in the inner part and the liquid one, also called "ocean", that formed on top of the solid crust due to the heat generated by the accretion. When the accretion has halted, the liquid ocean slowly becomes solid again as the temperature decreases. Since we see pulsations during our \textit{XMM-Newton} observation (assuming pulsations were also present during the \textit{Swift}/XRT observations), the heat stored in the crust must be preferably released at hot spots, likely at the magnetic poles. Therefore, it is plausible to assume that the magnetic field in the crust might be parallel to the surface and only exits the crust at the poles (Figure \ref{fig:Phase_model}).\\
\indent Once the cooling starts, the bottom parts of the ocean become solid. At this point, some magnetic field lines that were first in the liquid ocean are now in the new solid layer. It has been shown that in a solid crust layer the thermal conductivity is larger than when this layer is liquid (\citealt{Horowitz2007}; \citealt{Brown2009}; \citealt{Mckinven2016}). As a consequence, the heat flux toward the hot spots increases significantly in the layer that solidified. Therefore, we can distinguish two regions in the hot spots: one in the outer part that is produced by those field lines guiding heat from the liquid layers, and other in the central part of the hot spots that is produced by the magnetic field located in the solid crustal layer and shows a higher heat flux. If originally all the field lines were located in the liquid layers, then when part of those layers become solid, the inner part of the hot spot can suddenly rise in temperature. Thus the hot spot does not physically shrink in size, but its core becomes significantly hotter and the observed emission is dominated by this central region. Therefore we observe an artificially smaller size for the emitting region with higher temperature and luminosity. In addition, an extra amount of energy may be released due to the phase transition which would heat the inner part of the hot spot.\\
\indent This model could in principle explain what we observed in Interval IV, but we stress that detailed calculations should be performed to determine if this is also physically possible. In addition, one prediction of the model is that after the initial artificial observed decrease in size of the hot spot, it should appear to grow again when the cooling proceeds because most of the liquid layers will solidify and thus the region with enhanced heat flux will increase. We do not have additional observations after this enigmatic \textit{Swift}/XRT observation so we cannot test this prediction.\\

\subsection{Direct accretion onto neutron star magnetic poles}\label{subsec:accretion}
In the direct accretion scenario, the matter is guided by the magnetic field to the poles forming hot spots. Although it remains unknown how the accretion works at low accretion rate and what type of spectra would be produced, it might be possible that it would produce softer spectra than those obtained in higher accretion rate states (\citealt{Zampieri1995}). The Be/X-ray transient 1A 0535+26 was observed at similar low luminosities and it showed clear accretion features (source detection at $>$ 100 keV and aperiodic variability in the light curve; \citealt{Orlandini2004}, \citealt{Rothschild2013}; \citealt{Doroshenko2014}), so low-level accretion can definitely occur in these systems. However, its spectrum was harder than what we observe for 4U 0115+63 indicating that the emission is not formed in the same way in both systems. Two possibilities are left: either low-level accretion onto strongly magnetized NSs could give two types of spectra, or the physical mechanisms are widely different and we indeed could see cooling emission in 4U 0115+63.\\
\indent When assuming that the plateau phase in 4U 0115+63 is due to low-level accretion down to the surface of the NS, the behaviour seen during Interval IV could be due to some increase in accretion triggered by the periastron passage. However, if the direct accretion scenario is the right model to explain the source behaviour during this observation, there are still some caveats to be clarified. A temperature increase can naturally be explained if the accretion rate onto the pole increases, but it is not clear why the size of the emitting region would suddenly decrease significantly. This could signal a change in accretion geometry (e.g., disk like to wind like, or vice versa) but it is unclear if this is realistic. One key ingredient that needs to be explained is why we observe such a soft spectra in the plateau phase compared to the type-II and the mini-type I outbursts (\citealt{Wijnands2016}) where the source is definitely accreting. Detailed models are needed that calculate the expected spectral shape during these different accretion states in order to make any further conclusive statements.\\

\section{Conclusions}\label{sec:conclusions}
We have presented the latest results of our monitoring campaign using \textit{Swift} and \textit{XMM-Newton} on the Be/X-ray transient 4U 0115+63 during its low-luminosity state observed after its 2015 type-II outburst. The aim of our study was to investigate the possible physical mechanism or mechanisms behind this low-luminosity state that lasted for $\sim$200 days. The three previously proposed mechanisms  (\citealt{Wijnands2016}) to explain this state are: accretion down to the NS magnetosphere, accretion onto the magnetic poles of the NS,  or cooling emission from a NS heated due to the accretion of matter during the type-II outburst. The low number of photons detected during the low-luminosity state only allowed for simple, single component models (i.e., a power-law or blackbody models) to be fitted to the X-ray spectra. The spectra were relatively soft and were well described by the blackbody model. In addition, we detected pulsations in our \textit{XMM-Newton} observation, indicative of emission from the NS magnetic poles. As a consequence, the hypothesis of emission from the magnetospheric boundary is ruled out because pulsations are not expected in this scenario (assuming that pulsations were presented during the whole low-luminosity state; this could not be tested with the available data). However, we cannot conclusively reject the other two scenarios (or a combination of them), although both cannot straight forwardly explain our results. More dense monitoring of Be/X-ray transients after their type-II outbursts is needed to further determine the mechanism behind the low-luminosity state they can exhibit after those outbursts.\\

\section*{Acknowledgements}\label{acknowledgements}
ARE and RW acknowledge support from an NWO Top Grant, Module 1, awarded to RW. ASBN and AP are supported by an NWO Vidi grant, awarded to AP. YC is supported by the European Union Horizon 2020 research and innovation programme under the Marie Sklodowska-Curie Global Fellowship (grant agreement No.703916). ND acknowledges support from an NWO Vidi grant.\\



\begin{table*}
	\begin{center}
	\small
	\caption{Results of our spectral analysis using the blackbody model.}
	\label{tab:Table1}
	\begin{threeparttable}
	\begin{tabular}{ | r | c | c | c | c | c | c |}	
		\hline
		obsID & State & Exposure & kT$_{bb}$ & R$_{bb}$ & F$_{X}$ & L$_{X} $\\
		  &  & (ksec) & (keV) & (km) & (10$^{-12}$ erg cm$^{-2}$ s$^{-1}$) & (10$^{33}$ erg s$^{-1}$)\\
		\hline
		\smallskip
		0790180301 & XMM-Newton observation (b) & $\sim$20.8 & 0.44$\pm$0.01 & 0.44$^{+0.03}_{-0.02}$ & 0.15$\pm$0.01 & 0.90$\pm$0.03\\	
		\smallskip
		000311720+[42,45] & Interval III (a) & $\sim$5.8 & 0.47$^{+0.07}_{-0.06}$ & 0.51$^{+0.19}_{-0.13}$ & 0.27$^{+0.06}_{-0.05}$ & 1.6$\pm$0.3\\
		\smallskip
		+46 & Interval IV (c) & $\sim$7.9 & 0.85$^{+0.13}_{-0.11}$ & 0.18$^{+0.05}_{-0.04}$ & 0.35$^{+0.06}_{-0.05}$ & 2.1$^{+0.4}_{-0.3}$\\
		\smallskip
		+47 & 1st Type-I outburst (d) & $\sim$1.0 & 1.5$\pm$0.1 & 1.5$\pm$0.1 & 260$\pm$9 & 1525$^{+51}_{-50}$\\
		\smallskip		
		+48 & '' & $\sim$1.0 & 1.6$\pm$0.1 & 0.89$\pm$0.07 & 102$\pm$5 & 599$^{+27}_{-26}$\\
		\smallskip
		+49 & '' & $\sim$0.9 & 1.2$^{+0.2}_{-0.1}$ & 0.47$^{+0.10}_{-0.08}$ & 10$\pm$1 & 60$\pm$6\\
		\smallskip
		+50 & Quiescence I (e) & $\sim$8.8 & 0.36$^{+0.04}_{-0.05}$ & 0.3$^*$ & 0.03$\pm$0.02 & 0.18$^{+0.13}_{-0.09}$\\
		\smallskip
		+51 & 2nd Type-I outburst (f) & $\sim$1.0 & 1.0$^{+0.2}_{-0.1}$ & 0.58$^{+0.19}_{-0.15}$ & 7.5$^{+1.0}_{-0.9}$  & 44$^{+6}_{-5}$\\
		\smallskip
		+52 & '' & $\sim$0.8 & 1.8$\pm$0.2 & 0.41$\pm$0.06 & 32$\pm$2 & 186$\pm$14\\
		\smallskip	
		+53 & '' & $\sim$0.8 & 1.5$\pm$0.1 & 1.2$\pm$0.1 & 134$\pm$7 & 783$^{+42}_{-41}$\\
		\smallskip	
		+54 & '' & $\sim$1.0 & 1.5$\pm$0.1 & 1.5$\pm$0.1 & 201$\pm$8 & 1177$^{+45}_{-44}$\\
		\smallskip	
		+[55,59] & Quiescence II (g) & $\sim$4.7 & 0.49$\pm$0.05 & 0.3$^*$ & 0.12$^{+0.05}_{-0.04}$ & 0.68$^{+0.29}_{-0.23}$\\
		\hline
		\end{tabular}
	\begin{tablenotes}
	\scriptsize
	\item Notes: the unabsorbed X-ray fluxes and X-ray luminosities are given in the 0.5-10 keV energy range assuming a fixed N$_{H}\sim$ 9$\times$10$^{21}$ cm$^{-2}$ and a source distance of $\sim$7 kpc. The errors are 1$\sigma$. ($^{*}$)The fit parameter was fixed to this value.\\
	\end{tablenotes}
	\end{threeparttable}
	\end{center}
\end{table*}

\begin{table*}
	\begin{center}
	\small
	\caption{Results of our spectral analysis using the power-law model.}
	\label{tab:Table2}
	\begin{threeparttable}
	\begin{tabular}{ | r | c | c | c | c | c |}	
		\hline
		obsID & $\Gamma$ & F$_{X}$ & L$_{X}$\\
		  & & (10$^{-12}$ erg cm$^{-2}$ s$^{-1}$) & (10$^{33}$ erg s$^{-1}$)\\
		\hline
		\smallskip
		0790180301 & 2.6$\pm$0.1 & 0.26$\pm$0.01 & 1.5$\pm$0.1 \\	
		\smallskip
		000311720+[42,45] & 2.6$\pm$0.4 & 0.55$^{+0.10}_{-0.08}$ & 2.6$^{+0.6}_{-0.5}$ \\
		\smallskip
		+46 & 1.5$\pm$0.3 & 0.53$^{+0.10}_{-0.09}$ & 3.1$^{+0.6}_{-0.5}$ \\
		\smallskip
		+47 & 0.53$\pm$0.07 & 358$^{+17}_{-16}$ & 2096$^{+98}_{-93}$ \\
		\smallskip		
		+48 & 0.59$\pm$0.09 & 130.1$^{+7.6}_{-7.2}$ & 762$^{+45}_{-42}$ \\
		\smallskip
		+49 & 1.2$\pm$0.3 & 15$\pm$2 & 86$^{+10}_{-9}$ \\
		\smallskip
		+50 & 3.6$^{+3.0}_{-1.8}$ & 0.08$^{+0.52}_{-0.05}$ & 0.47$^{+3.03}_{-0.30}$\\
		\smallskip
		+51 & 1.6$\pm$0.5 & 13$\pm$2 & 73$^{+12}_{-9}$ \\
		\smallskip
		+52 & 0.60$\pm$0.19 & 39$^{+4}_{-3}$ & 230$^{+22}_{-20}$ \\
		\smallskip	
		+53 & 0.70$\pm$0.12 & 181$^{+13}_{-12}$ & 1058$^{+75}_{-70}$ \\
		\smallskip	
		+54 & 0.69$\pm$0.08 & 270$^{+14}_{-13}$ & 1581$^{+80}_{-76}$ \\
		\smallskip	
		+[55,59] & 2.1$\pm$1.4 & 0.22$^{+0.29}_{-0.08}$ & 1.3$^{+1.7}_{-0.4}$ \\
		\hline
	\end{tabular}
	\begin{tablenotes}
	\scriptsize
	\item Notes: the unabsorbed X-ray fluxes and X-ray luminosities are given in the 0.5-10 keV energy range assuming a fixed N$_{H}\sim$ 9$\times$10$^{21}$ cm$^{-2}$ and a source distance of $\sim$7 kpc. The errors are 1$\sigma$. ($^{*}$)The fit parameter was fixed to this value.\\
	\end{tablenotes}
	\end{threeparttable}
	\end{center}
\end{table*}

\bsp	
\label{lastpage}
\end{document}